\documentclass[fleqn]{an}
\usepackage{graphicx}
\usepackage{times}
\usepackage{bm}
\Pagespan{43}{50}
\Yearpublication{2011}
\Yearsubmission{2010}
\Month{9}
\Volume{332}
\Issue{1}
\DOI{10.1002/asna.201012345}
\begin{document}
\sloppy

\newcommand{\EQ}{\begin{equation}}
\newcommand{\EE}{\end{equation}}
\newcommand{\EQA}{\begin{eqnarray}}
\newcommand{\EEA}{\end{eqnarray}}
\newcommand{\brac}[1]{\langle #1 \rangle}
\newcommand{\pd}{\partial}
\newcommand{\pdz}{\partial_z}
\newcommand{\DIV}{\vec{\nabla} \cdot }
\newcommand{\CURL}{\vec{\nabla} \times }
\newcommand{\cross}[2]{\boldsymbol{#1} \times \boldsymbol{#2}}
\newcommand{\crossm}[2]{\brac{\boldsymbol{#1}} \times \brac{\boldsymbol{#2}}}
\newcommand{\ve}[1]{\boldsymbol{#1}}
\newcommand{\mean}[1]{\overline{#1}}
\newcommand{\meanv}[1]{\overline{\bm #1}}
\newcommand{\cst}{c_{\rm s}^2}
\newcommand{\nut}{\nu_{\rm t}}
\newcommand{\etat}{\eta_{\rm t}}
\newcommand{\etatz}{\eta_{\rm t0}}
\newcommand{\chit}{\chi_{\rm t}}
\newcommand{\memf}{\overline{\bm{\mathcal{E}}}}
\newcommand{\memfi}{\overline{\mathcal{E}}_i}
\newcommand{\etaT}{\eta_{\rm T}}
\newcommand{\urms}{u_{\rm rms}}
\newcommand{\Urms}{U_{\rm rms}}
\newcommand{\brms}{B_{\rm rms}}
\newcommand{\Beq}{B_{\rm eq}}
\newcommand{\eu}{\hat{\bm e}}
\newcommand{\xu}{\hat{\bm x}}
\newcommand{\yu}{\hat{\bm y}}
\newcommand{\zu}{\hat{\bm z}}
\newcommand{\Ou}{\hat{\bm \Omega}}
\newcommand{\kef}{k_{\rm f}}
\newcommand{\tauc}{\tau_{\rm c}}
\newcommand{\tauto}{\tau_{\rm to}}
\newcommand{\HP}{H_{\rm P}}
\newcommand{\St}{{\rm St}}
\newcommand{\Sh}{{\rm Sh}}
\newcommand{\Pm}{{\rm Pm}}
\newcommand{\Rm}{{\rm Rm}}
\newcommand{\Pra}{{\rm Pr}}
\newcommand{\Ra}{{\rm Ra}}
\newcommand{\Ma}{{\rm Ma}}
\newcommand{\Tay}{{\rm Ta}}
\newcommand{\Ro}{{\rm Ro}}
\newcommand{\Rey}{{\rm Re}}
\newcommand{\Co}{{\rm Co}}
\newcommand{\Cost}{\Omega_\star}
\newcommand{\ReLS}{{\rm Re}_{\rm LS}}
\newcommand{\qxx}{Q_{xx}}
\newcommand{\qyy}{Q_{yy}}
\newcommand{\qzz}{Q_{zz}}
\newcommand{\qxy}{Q_{xy}}
\newcommand{\qxz}{Q_{xz}}
\newcommand{\qyz}{Q_{yz}}
\newcommand{\qij}{Q_{ij}}
\newcommand{\Omx}{\Omega_x}
\newcommand{\Omz}{\Omega_z}
\newcommand{\emf}{\bm{\mathcal{E}}}
\newcommand{\emfi}{\mathcal{E}_i}
\newcommand{\nab}{\mbox{\boldmath $\nabla$} {}}
\newcommand{\meanFFFF}{\overline{\mbox{\boldmath ${\cal F}$}}{}}{}
\def\onethird{{\textstyle{1\over3}}}
\def\onehalf{{\textstyle{1\over2}}}
\def\threefourths{{\textstyle{3\over4}}}
\def\threehalfs{{\textstyle{3\over2}}}

\title{Effects of stratification in spherical shell convection}

\author{P.J. K\"apyl\"a\inst{1,2}\fnmsep\thanks{Corresponding author:
    {petri.kapyla@helsinki.fi}}, M.J. Mantere\inst{1}, and A. 
    Brandenburg\inst{2,3}}

\titlerunning{Effects of stratification in spherical shell convection}
\authorrunning{P.J. K\"apyl\"a et al.}

\institute{ 
Department of Physics, PO BOX 64 (Gustaf H\"allstr\"omin katu 2a), 
FI-00014 University of Helsinki, Finland
\and
Nordita\thanks{Nordita is a Nordic research institute jointly operated
by the Stockholm University and the Royal Institute of Technology, Stockholm.},
AlbaNova University Center, Roslagstullsbacken 23, SE-10691 
Stockholm, Sweden
\and
Department of Astronomy, Stockholm University, SE-10691 
Stockholm, Sweden}

\received{2011 Aug 31} 
\accepted{2011 Nov 10}
\publonline{2012 Jan 12}


\abstract{%
  We report on simulations of mildly turbulent convection in spherical
  wedge geometry with varying density stratification. We vary the
  density contrast within the convection zone by a factor of 20 and
  study the influence of rotation on the solutions. We demonstrate that the
  size of convective cells decreases and the anisotropy of turbulence
  increases as the stratification is increased. Differential rotation
  is found to change from anti-solar (slow equator) to solar-like (fast equator)
  at roughly the same Coriolis number for all stratifications. The
  largest stratification runs, however, are sensitive to changes of
  the Reynolds number.
  Evidence for a near-surface shear layer is found in runs with
  strong stratification and large Reynolds numbers.
}
\maketitle

\section{Introduction}
\label{sec:intro}

Numerical simulations of turbulent convection in spherical geometry
have become a standard tool in the study of differential rotation and
magnetism in the solar and stellar context (Miesch \& Toomre
2009). The current state of the art models can reproduce many aspects
of the solar internal rotation (e.g.\ Miesch et al.\ 2006), and
large-scale oscillatory dynamo action occurs when rotation is rapid
enough (Brown et al.\ 2010, 2011). However, reproducing the solar
cycle has turned out to be elusive, even though large-scale oscillatory
fields are now seen in some simulations with the solar rotation rate, too
(e.g.\ Ghizaru et al.\ 2010; Racine et al.\ 2011).
The reason for the remaining discrepancies may lie in the fact that
much of the physics have either to be simplified or neglected
altogether due to severe numerical constraints (e.g.\ K\"apyl\"a
2011). Furthermore, the simulations that are being carried out are so demanding
that often only a single or a few representative cases can be done.
Although many results, such as the change from anti-solar (slow
equator) to solar-like (fast equator) differential rotation as the
rotation rate increases (e.g.\ Chan 2010; K\"apyl\"a et al.\ 2011a),
and the appearance of mostly axisymmetric large-scale magnetic fields
(e.g.\ Gilman 1983; Glatzmaier 1985, Browning et al.\ 2006; 
Brown et al.\ 2010, 2011;
K\"apyl\"a et al.\ 2010) appear robust, their exact dependence on
different simulation parameters has not been explored in detail.

Here we study the effect of density stratification on rotating spherical shell
convection. Our main goal is to study how the transition from
anti-solar to solar-like rotation is affected. This is relevant
because most convection models in spherical shells still have rather
modest density stratification in comparison to the Sun. Although most
of the mass within the convection zone is located near the base,
fast downflows at the vertices of convection cells originate near the
surface.
The effect of these downflows on angular momentum transport is yet
unclear.
We are also interested in the statistical properties, such as
anisotropy, of turbulence as the stratification is increased.

\section{Model}
\label{sec:model}

Our model is based on that used by K\"apyl\"a et al.\ (2010, 2011a). We
model a segment of a star, i.e.\ a ``wedge'', in spherical polar
coordinates, where $(r,\theta,\phi)$ denote the radius, colatitude, and
longitude. The radial, latitudinal, and longitudinal extents of the
computational domain are given by $0.7R \leq r \leq R$, $\theta_0 \leq
\theta \leq \pi-\theta_0$, and $0 \leq \phi \leq \phi_0$,
respectively, where $R$ is the radius of the star. In all of our runs
we take $\theta_0=\pi/8$ and $\phi_0=\pi/2$.

We solve the following equations of compressible hydrodynamics
in a frame of reference rotating with angular velocity $\bm\Omega_0$,
\begin{equation}
\frac{D \ln \rho}{Dt} = -\bm\nabla\cdot\bm{u},
\end{equation}
\begin{equation}
\frac{D\bm{u}}{Dt} = \bm{g} -2\bm\Omega_0\times\bm{u}+\frac{1}{\rho}
\left(\bm\nabla \cdot 2\nu\rho\bm{\mathsf{S}}-\bm\nabla p\right),
\end{equation}
\begin{equation}
T\frac{D s}{Dt} = \frac{1}{\rho}\left[\bm\nabla \cdot (K \bm\nabla T)
+ \bm\nabla \cdot (\rho T \chi_{\rm t} \bm\nabla s) + 2\nu \bm{\mathsf{S}}^2\right],
\label{equ:ss}
\end{equation}
where $D/Dt = \pd/\pd t + \bm{u} \cdot \bm\nabla$ is the advective
time derivative, $\rho$ is the density, $\bm{u}$ is the velocity, $s$
is the specific entropy, $T$ is the temperature, and $p$ is the
pressure. The fluid obeys the ideal gas law with $p=(\gamma-1)\rho e$,
where $\gamma=c_{\rm P}/c_{\rm V}=5/3$ is the ratio of specific heats
at constant pressure and volume, respectively, and $e=c_{\rm V} T$ is
the internal energy.

Furthermore, $\nu$ is the kinematic viscosity, $K$ is the radiative
heat conductivity, $\chi_{\rm t}$ is the unresolved turbulent heat
conductivity, and $\bm{g}$ is the gravitational acceleration given by
\begin{equation}
\bm{g}=-\frac{GM}{r^2}\hat{\bm{r}},
\end{equation}
where $G$ is the gravitational constant, $M$ is the mass of the star,
and $\hat{\bm{r}}$ is the unit vector in the radial direction. We omit
the centrifugal force in our models. The rate of strain tensor
$\bm{\mathsf{S}}$ is given by
\begin{equation}
\mathsf{S}_{ij}=\onehalf(u_{i;j}+u_{j;i})-\onethird \delta_{ij}\bm\nabla\cdot\bm{u},
\end{equation}
where the semicolons denote covariant differentiation; see Mitra et
al.\ (2009) for details. Unlike in our previous studies
(K\"apyl\"a et al.\ 2010, 2011a), we omit
stably stratified layers below and above the convectively unstable
layer.

\subsection{Initial and boundary conditions}
\label{sec:initcond}
\begin{figure}
\resizebox{\hsize}{!}
{\includegraphics{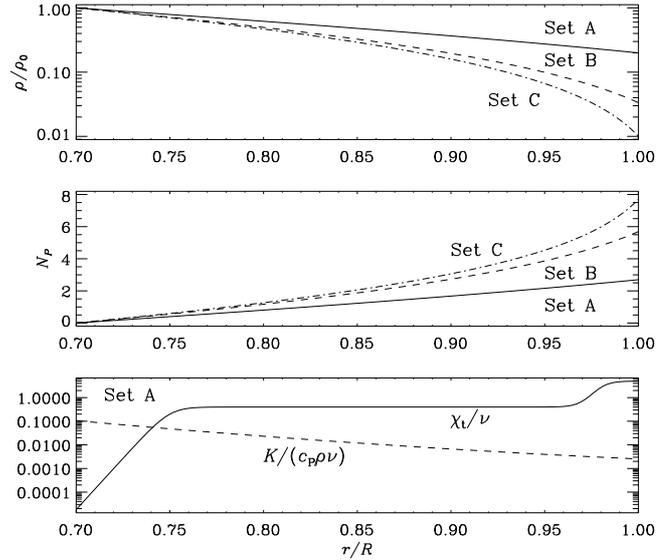}}
\caption{Profiles of density (top panel), number of pressure scale
  heights, $N_{\rm P}=\ln\frac{p_0}{p}$, where $p_0$ is the pressure 
  at $r=0.7R$ (middle), and $\chit$ and $K$
  (bottom). In the bottom panel $\nu$ and $\rho$ from Set~A are used
  as normalization factors.}
\label{piprofs}
\end{figure}

In the initial state the atmosphere is adiabatic and the hydrostatic
temperature gradient is given by
\begin{equation}
\frac{\pd T}{\pd r} = \frac{-g}{c_{\rm V}(\gamma-1)(m+1)},
\label{eq:dTdz}
\end{equation}
where $m=1.5$ is the polytropic index. We use Eq.~(\ref{eq:dTdz}) as
the lower boundary condition for the temperature. This gives the
logarithmic temperature gradient $\nabla$ (not to be confused with the
operator $\bm\nabla$) as
\begin{equation}
\nabla=\pd \ln T/\pd \ln p = (m+1)^{-1}.
\end{equation}
Density stratification is obtained by requiring hydrostatic
equilibrium. The heat conduction profile is chosen so that radiative
diffusion is responsible for supplying the energy flux in the system,
and $K$ decreases rapidly within the convection zone (see,
Fig.~\ref{piprofs}).

The radial and latitudinal boundaries are taken to be impenetrable and
stress free, according to
\begin{eqnarray}
\lefteqn{u_r=0,\quad \frac{\pd u_\theta}{\pd r}=\frac{u_\theta}{r},\quad \frac{\pd
u_\phi}{\pd r}=\frac{u_\phi}{r} \quad (r=0.7R, R),}\\
\lefteqn{\frac{\pd u_r}{\pd \theta}=u_\theta=0,\quad \frac{\pd u_\phi}{\pd
\theta}=u_\phi \cot \theta \quad (\theta=\theta_0,\pi-\theta_0).}
\end{eqnarray}
On the latitudinal boundaries we assume that the thermodynamic
quantities have zero first derivative, thus suppressing heat fluxes
through the boundaries.

On the upper boundary we apply a black body condition
\begin{equation}
\sigma T^4  = -K\frac{\pd T}{\pd r} - \rho T \chi_{\rm t} \frac{\pd s}{\pd r},
\end{equation}
where $\sigma$ is the Stefan--Boltzmann constant. In our runs
we use a modified value for $\sigma$ that takes into account that our
Reynolds and Rayleigh numbers are much smaller than in reality,
so $K$ is much larger and therefore the flux too high.
The black body boundary for the temperature has previously
been used in mean-field models of Brandenburg et al.\ (1992).
In our runs $K$ is negligibly small near the surface so that the
unresolved convective energy flux transports practically all of the
energy through the upper boundary. This is similar to what is commonly
used in the ASH simulations (e.g.\ Brun et al.\ 2004).

\subsection{Dimensionless parameters}
We obtain non-dimensional quantities by choosing
\begin{eqnarray}
R = GM = \rho_0 = c_{\rm P} = 1\;,
\end{eqnarray}
where $\rho_0$ is the density at $0.7R$. The units of length,
velocity, density, and entropy are then given by
\begin{eqnarray}
\lefteqn{ [x] = R\;,\;\; [u]=\sqrt{GM/R}\;,\;\;[\rho]=\rho_0\;,\;\;[s]=c_{\rm P}\;.}
\end{eqnarray}
The simulations are governed by the Prandtl, Reynolds, Coriolis, and
Rayleigh numbers, defined by
\begin{eqnarray}
\Pra&\!=\!&\frac{\nu}{\chit}\;,\;\;\Rey=\frac{\urms}{\nu \kef}\;,\;\;
\Co=\frac{2\Omega_0}{\urms \kef}\;,\\
\Ra&\!=\!&\frac{GM(\Delta r)^4}{\nu \chit R^2} \bigg(-\frac{1}{c_{\rm P}}\frac{{\rm d}s}{{\rm d}r} \bigg)_{r_{\rm m}}\;,
\label{equ:Co}
\end{eqnarray}
where $\chit$ is the turbulent thermal conductivity in the middle of
the convection zone (i.e.\ at $r_{\rm m}=0.85R$), $\kef=2\pi/\Delta r$
is an estimate of the wavenumber of
the energy-carrying eddies, $\Delta r=0.3R$ is the thickness of the
layer, and
$\urms=\sqrt{\threehalfs\brac{u_r^2+u_\theta^2}}$ is the rms velocity,
where the angle brackets denote volume averaging. In our definition
of $\urms$ we omit the contribution from the $\phi$-velocity,
because its value is dominated by effects from the differential rotation
(K\"apyl\"a et al.\ 2011a). Sometimes we show $\urms(r)$ which is
the fluctuating rms velocity as a function of radius and from which
we have subtracted the azimuthally averaged velocities.
The entropy gradient, measured at $r_{\rm m}$, is given by
\begin{eqnarray}
\bigg(-\frac{1}{c_{\rm P}}\frac{{\rm d}s}{{\rm d}r}\bigg)_{r_{\rm m}} = \frac{\nabla_{\rm m}-\nabla_{\rm ad}}{H_{\rm P}}\;,
\end{eqnarray}
where $\nabla_{\rm m} = (\pd \ln T/\pd \ln p)_{r_{\rm m}}$, 
and $H_{\rm P}$ is the pressure scale height at $r_{\rm m}$.
Due to the fact that the initial stratification is isentropic, we
quote values of $\Ra$ from the thermally saturated state of the runs.

The energy that is deposited into the domain at the base is
controlled by the luminosity parameter
\begin{equation}
\mathcal{L} = \frac{L_0}{\rho_0 (GM)^{3/2} R^{1/2}},
\end{equation}
where $L_0=4\pi r_1^2 F_{\rm b}$ is the constant luminosity, and
$F_{\rm b}=-(K \pd T/\pd r)|_{r=0.7R}$ is the energy flux imposed at
the bottom boundary. Furthermore, the stratification is determined by
the normalized pressure scale height at the surface
\begin{eqnarray}
\xi = \frac{(\gamma-1) c_{\rm V}T_1}{GM/R},
\end{eqnarray}
where $T_1=T(r=R)$. Similar parameter definitions were used by
Dobler et al.\ (2006). We use three different values 
$(0.09,0.02,8\cdot10^{-3})$ of $\xi$ which
result in density contrasts of 5, 30, and $10^2$, respectively
(see Fig.~\ref{piprofs}). Now the convection zones span between
roughly 2.5 and 7.5 pressure scale heights.

The simulations were performed using the {\sc Pencil Code}%
\footnote{\texttt{http://pencil-code.googlecode.com/}}, which uses
sixth-order explicit finite differences in space and a third-order
accurate time stepping method; see Mitra et al.\ (2009) for further
information regarding the adaptation of the {\sc Pencil Code} to
spherical coordinates.

\begin{table*}
 \centering
 \caption{Summary of the runs. Here, ${\rm Ma}=\urms/\sqrt{GM/R}$, 
   $k_\Omega=\Delta \Omega/\Omega_{\rm eq}$,
   $\Delta \Omega = \Omega_{\rm eq}-\Omega_{\rm pole}$, where 
   $\Omega_{\rm eq}=\mean{\Omega}(R,\theta=\pi/2)$ and 
   $\Omega_{\rm pole}=\mean{\Omega}(R,\theta=\theta_0)$. Furthermore, 
   $E_{\rm k}=\brac{\onehalf \rho \bm{u}^2}$ is the volume 
   averaged total kinetic energy, and
   $E_{\rm m}=\onehalf\brac{\rho(\mean{u}_r^2+\mean{u}_\theta^2)}$ 
   and $E_{\rm r}=\onehalf\brac{\rho\mean{u}_\phi^2}$ are the kinetic
   energies of the meridional circulation and differential rotation,
   respectively.
   The coefficients $\omega_1$, $\omega_3$, and $\omega_5$ represent the
   expansion coefficients in Eq.~(\ref{equ:Omfit}) and will be discussed
   in Sect.~\ref{sec:diffrot} below.
}\label{tab:runs}
\begin{tabular}{lcccccccccrrrr}\hline
Run  & $\xi$ & $\Ra$ & $\Pr$ & $\mathcal{L}$ & $\Ma$ & $\Rey$ & $\Co$ & $E_{\rm m}/E_{\rm k}$ & $E_{\rm r}/E_{\rm k}$ & $k_\Omega$
& $\omega_1$ & $\omega_3$ & $\omega_5$ \\
\hline
A0 &     $0.09$     & $1.5\cdot10^5$ & 2.5 & $3.8\cdot10^{-5}$ & 0.025 & 41 & ---- & 0.116 & ----  & ----    & ---- & ----    & ----    \\ 
A1 &     $0.09$     & $5.7\cdot10^5$ & 2.5 & $3.8\cdot10^{-5}$ & 0.020 & 33 & 1.37 & 0.003 & 0.944 & $-2.71$ & 0.78 & $ 0.15$ & $ 0.03$ \\ 
A2 &     $0.09$     & $1.2\cdot10^6$ & 2.5 & $3.8\cdot10^{-5}$ & 0.016 & 26 & 3.48 & 0.000 & 0.798 & $-0.13$ & 1.06 & $-0.04$ & $ 0.02$ \\ 
A3 &     $0.09$     & $1.7\cdot10^6$ & 2.5 & $3.8\cdot10^{-5}$ & 0.014 & 23 & 5.97 & 0.000 & 0.889 & $0.19$  & 1.06 & $-0.05$ & $ 0.01$ \\ 
A4 &     $0.09$     & $2.2\cdot10^6$ & 2.5 & $3.8\cdot10^{-5}$ & 0.013 & 21 & 8.78 & 0.000 & 0.914 & $0.15$  & 1.05 & $-0.04$ & $ 0.01$ \\ 
\hline
B0 &     $0.02$     & $2.4\cdot10^5$ &  5  & $3.8\cdot10^{-5}$ & 0.027 & 22 & ---- & 0.037 & ----  & ----    & ---- & ----    & ----    \\ 
B1 &     $0.02$     & $3.2\cdot10^5$ &  5  & $3.8\cdot10^{-5}$ & 0.026 & 22 & 1.04 & 0.005 & 0.906 & $-2.12$ & 0.79 & $ 0.12$ & $ 0.01$ \\ 
B2 &     $0.02$     & $5.3\cdot10^5$ &  5  & $3.8\cdot10^{-5}$ & 0.025 & 20 & 2.24 & 0.004 & 0.879 & $-0.59$ & 1.06 & $ 0.09$ & $-0.02$ \\ 
B3 &     $0.02$     & $3.0\cdot10^6$ & 2.5 & $3.8\cdot10^{-5}$ & 0.024 & 40 & 4.54 & 0.001 & 0.786 & $0.03$  & 1.06 & $-0.01$ & $ 0.01$ \\ 
B4 &     $0.02$     & $3.2\cdot10^6$ & 2.5 & $3.8\cdot10^{-5}$ & 0.021 & 36 & 7.61 & 0.001 & 0.581 & $0.08$  & 1.05 & $-0.02$ & $ 0.01$ \\ 
\hline
C0 & $8\cdot10^{-3}$ & $3.4\cdot10^5$ &  5  & $6.3\cdot10^{-6}$ & 0.017 & 26 & ---- & 0.075 & ----  & ----    & ---- & ----    & ----     \\ 
C1 & $8\cdot10^{-3}$ & $4.8\cdot10^5$ &  5  & $6.3\cdot10^{-6}$ & 0.017 & 25 & 0.91 & 0.003 & 0.772 & $-0.97$ & 0.92 & $ 0.05$ & $ 0.02$ \\ 
C2 & $8\cdot10^{-3}$ & $8.2\cdot10^5$ &  5  & $6.3\cdot10^{-6}$ & 0.016 & 25 & 1.85 & 0.002 & 0.760 & $-0.43$ & 0.90 & $ 0.05$ & $-0.01$ \\ 
C3 & $8\cdot10^{-3}$ & $1.7\cdot10^5$ &  5  & $6.3\cdot10^{-6}$ & 0.015 & 22 & 4.12 & 0.003 & 0.333 & $-0.07$ & 1.00 & $-0.00$ & $ 0.03$ \\ 
C4 & $8\cdot10^{-3}$ & $2.1\cdot10^5$ &  5  & $6.3\cdot10^{-6}$ & 0.013 & 19 & 7.22 & 0.003 & 0.250 & $-0.01$ & 1.00 & $-0.00$ & $ 0.03$ \\ 
\hline
D0 & $8\cdot10^{-3}$ & $2.8\cdot10^5$ &  2  & $6.3\cdot10^{-6}$ & 0.019 & 73 & ---- & 0.078 & ----  & ----    & ---- & ----    & ----     \\ 
D1 & $8\cdot10^{-3}$ & $5.5\cdot10^5$ &  2  & $6.3\cdot10^{-6}$ & 0.019 & 72 & 0.79 & 0.002 & 0.826 & $-1.50$ & 0.94 & $ 0.06$ & $ 0.03$ \\ 
D2 & $8\cdot10^{-3}$ & $1.3\cdot10^6$ &  2  & $6.3\cdot10^{-6}$ & 0.018 & 66 & 1.72 & 0.002 & 0.956 & $-1.52$ & 0.81 & $ 0.11$ & $ 0.00$ \\ 
D3 & $8\cdot10^{-3}$ & $3.1\cdot10^5$ &  2  & $6.3\cdot10^{-6}$ & 0.018 & 68 & 3.35 & 0.002 & 0.589 & $-0.14$ & 0.99 & $ 0.00$ & $ 0.00$ \\ 
D4 & $8\cdot10^{-3}$ & $1.8\cdot10^5$ & 2.5 & $6.3\cdot10^{-6}$ & 0.018 & 54 & 5.11 & 0.001 & 0.718 & $-0.00$ & 1.02 & $-0.02$ & $ 0.01$ \\ 
\hline
\end{tabular}
\end{table*}

\begin{figure}
\resizebox{\hsize}{!}
{\includegraphics{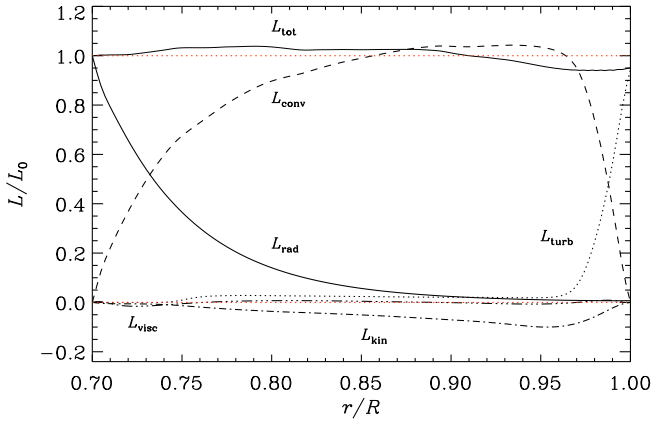}}
\caption{Flux balance from Run~B3. The different contributions are due
  to radiative diffusion (solid line), resolved convection (dashed),
  unresolved turbulence (dotted), flux of kinetic energy (dot-dashed),
  and viscosity (triple-dot-dashed). The red dotted lines denote the
  zero level and the total luminosity through the lower boundary.}
\label{pflux_B3}
\end{figure}

\begin{figure*}
\resizebox{\hsize}{!}
{\includegraphics{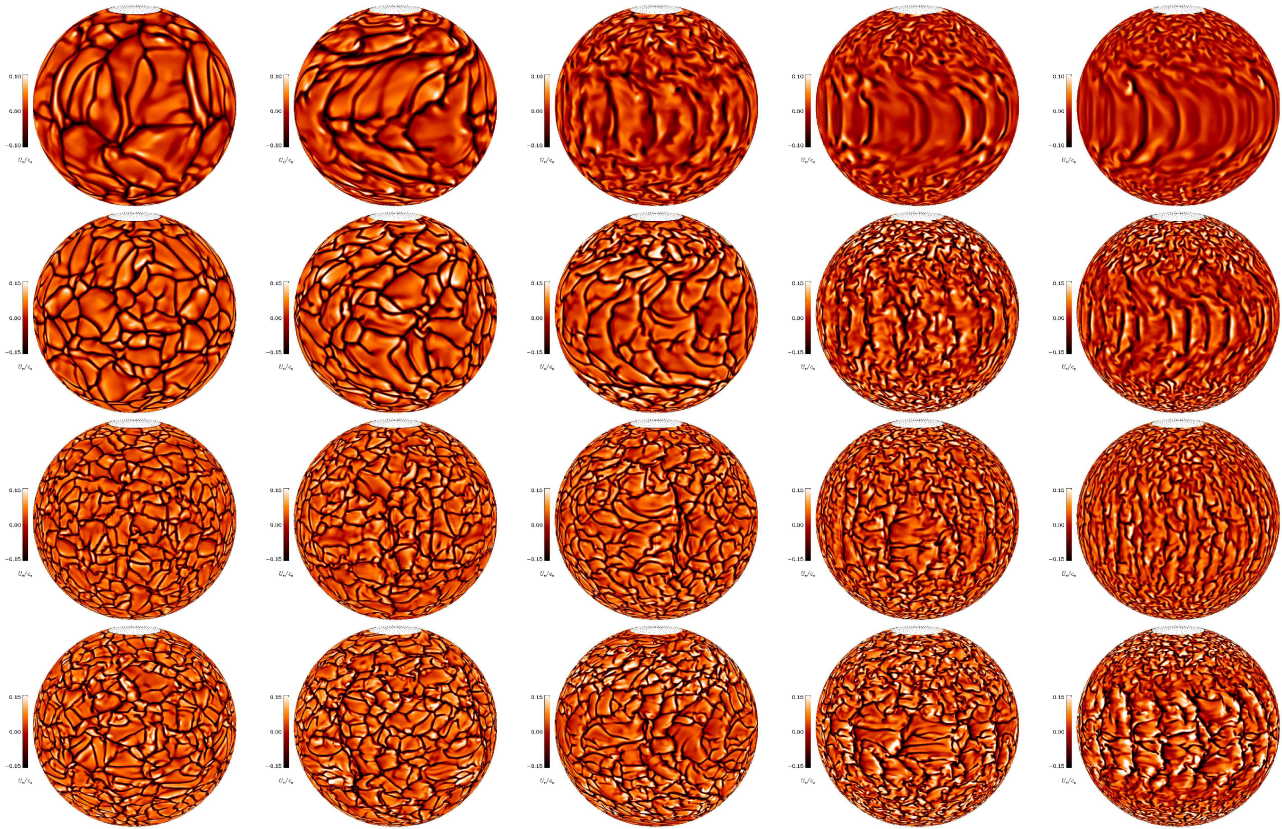}}
\caption{Radial velocity $u_r$, normalized by the local sound speed
  from $r=0.98R$ for runs in Sets~A (top row), B, C, and D
  (bottom row). The rotation rate increases from left to right. The
  longitude extent has been duplicated fourfold for visualization
  purposes.}
\label{pballs}
\end{figure*}

\section{Results}
\label{sec:results}

We have performed four sets of simulations differing by their density
stratification (see Table~\ref{tab:runs}) and Reynolds number. We vary 
the rotation rate
within each set so that the Coriolis number changes by roughly an order
of magnitude. We increase the gravity by a factor of ${10\over 3}$ in
Sets~C and D in order to limit the Mach number to roughly 0.1 near the
surface. The grid resolution in Sets~A--C is
$128\times256\times128$. In Set~C this means that the ratio of the
pressure scale height to the radial grid spacing at the surface is
$H_{\rm P}/\delta r\approx3.4$, which is on the limit of resolving the
structure.
We have remeshed snapshots from the saturated states of the runs in
Set~C to double resolution (Set~D) where we are also able to increase
the Reynolds number.

\subsection{Flux balance}
In contrast to our earlier studies using a polytropic setup with
$m=1$ (K\"apyl\"a et al.\ 2010, 2011a), we now
use a setup in which convection transports the majority of the flux.
This is achieved by decreasing the heat conductivity $K$ within the
convection zone and introducing a turbulent heat conductivity
$\chi_{\rm t}$ which is responsible for unresolved convective
transport of heat (e.g.\ Chan \& Sofia 1996; Brun et al.\ 2004). We
apply a constant value of $\chi_{\rm t}$, of the order of the kinematic
viscosity $\nu$, in the bulk of the convection zone ($0.75R<r<0.98R$)
and an order of magnitude larger value above $r>0.98R$ in order to
transport the flux through the upper boundary.
Below $r=0.75R$, $\chi_{\rm t}$ goes smoothly to zero, see
Fig.~\ref{piprofs}.

To verify that the system is in thermal equilibrium, we consider the
radiative, convective, kinetic, viscous, and turbulent energy fluxes,
defined as
\begin{eqnarray}
F_{\rm rad} &=& - K\frac{\pd \mean{T}}{\pd r}, \\
F_{\rm conv} &=& - c_{\rm P}\mean{\rho} \mean{u_r' T'}, \\
F_{\rm kin} &=& \onehalf \mean{\rho} \mean{u^2 u_r}, \\
F_{\rm visc} &=& -2\nu \mean{\rho}\ \mean{u_i\mathsf{S}_{ir}}, \\
F_{\rm turb} &=& - \mean{\rho} \mean{T} \chi_{\rm t} \frac{\pd \mean{s}}{\pd r},
\end{eqnarray}
where the averages are taken over $\theta$ and $\phi$. Representative
results from Run~B3 are shown in Fig.~\ref{pflux_B3}. Radiative
diffusion transports the total flux through the lower boundary and
decreases rapidly as a function of $r$. The radiative flux is less
than 10 per cent above $r=0.85R$. The flux due to resolved convection
is responsible for transporting the majority of the luminosity within
the convection zone. The flux of kinetic energy is directed downwards
and is responsible for roughly 10 per cent of the flux near the
surface. Note that the maxima of $F_{\rm conv}$ and $F_{\rm kin}$ are
significantly larger in the non-rotating cases. The unresolved
turbulent flux is small in the bulk of the convection zone and carries
the flux out through the outer boundary. The viscous flux is small in
all of our runs.
The flux balance is similar to the ASH simulations (e.g.\ Brun et al.\
2004; Miesch et al.\ 2008) which employ stratification from a 1D solar
model and somewhat different profiles of the diffusion coefficients.

\subsection{Properties of convection}
Visualizations of the radial velocity near the surface of the star for
all of our models are shown in Fig.~\ref{pballs}.
We find that as the stratification increases the size of convection cells
decreases. This is a consequence of the decreasing pressure scale
height near the surface.
In the case of the highest stratification, Sets~C and D, the granulation
pattern is similar to the high-resolution run reported by Miesch et
al.\ (2008) with a comparable stratification.
As noted above, the resolution in Set~C is close to critical in
resolving the stratification properly which is also manifested by
numerical artefacts in Fig.~\ref{pballs}. The higher resolution runs
in Set~D, however, are well behaved and show similar convection
patterns in the non-rotating and slowly rotating cases.

\begin{figure}
\resizebox{\hsize}{!}
{\includegraphics{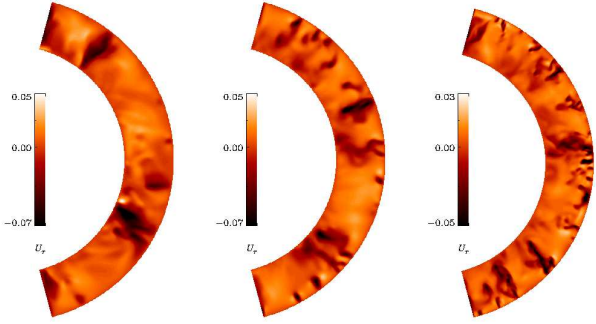}}
\caption{Radial velocity $u_r$ in the meridional plane $\phi=\phi_0$ 
from Runs~A0, 
B0, and D0 in units of $\sqrt{GM/R}$.}
\label{meri_Ur}
\end{figure}

Furthermore, as the rotation rate is increased, one sees the
formation of a cartridge belt-like pattern that is also known as
banana cells.
However, these structures become less pronounced at larger
stratification in Set~C. In Set~D, on the other hand, strong banana
cells are again observed. The main difference between Sets~C and D is
that in the latter the Reynolds number is modestly increased. 
The smaller size
of convection cells at high stratification leads to a smaller
``effective'' Reynolds number based on the horizontal eddy size,
which is not well reflected by our
definition of $\Rey$. It is possible that in Run~C4 this effective
$\Rey$ is below critical to excite the formation of banana cells and
strong prograde differential rotation.

Another way to see the difference between weak and strong
stratification is to visualize the flows in the meridional plane. In
Fig.~\ref{meri_Ur} we show a cut of the radial velocity at
$\phi=\phi_0$ from Runs~A0, B0, and D0 in the saturated
state. Whereas the downflows in Run~A0 go easily through the whole
convection zone, smaller-scale structures originating from the surface
appear already in Run~B0. For the strongest stratification the
anisotropy of the flow is clear also to the naked eye with smooth
large-scale flows in the deep layers and small-scale irregular flows
near the surface.
However, the strongest downflows are still able to go all the way
through the convection zone.

\begin{figure}
\resizebox{\hsize}{!}
{\includegraphics{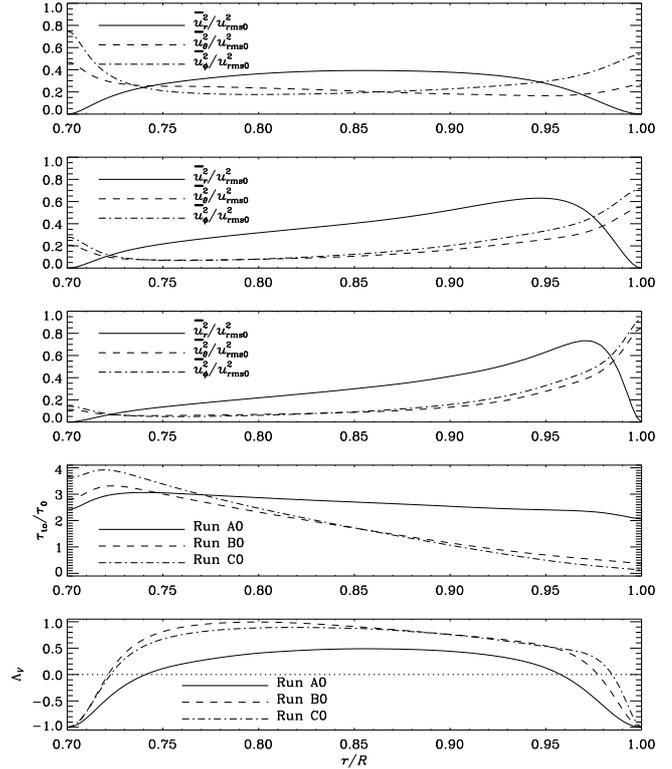}}
\caption{From top to bottom: the three uppermost panels show the
  radial dependence of the fluctuating velocity components averaged
  over $\theta$ and $\phi$ for nonrotating Runs~A0, B0, and C0,
  respectively. The fourth and fifth panel show the turnover time 
  $\tau_{\rm to}=H_{\rm P}/\urms(r)$ divided by $\tau_0=(u_{\rm rms0} \kef)^{-1}$, 
  and the vertical anisotropy parameter
  $\Lambda_{\rm V}$ for the same runs as indicated by the
  legends. Here $u_{\rm
    rms0}=\sqrt{\threehalfs\brac{u_r^2+u_\theta^2}}$ is used as a
    normalization factor.}
\label{puur}
\end{figure}

The squares of the fluctuating velocity components are shown in
Fig.~\ref{puur}. For the weakest stratification the profiles are
almost symmetrical with respect to $r_{\rm m}$. The apparent asymmetry
of the $\theta$ and $\phi$ velocities is likely due to the large size
of the convective cells in comparison to the domain size. For larger
stratification the velocities increase near the surface and the
anisotropy of the horizontal velocities decreases. An important effect
that follows is that the convective turnover time, defined as
\begin{equation}
\tau_{\rm to}=H_{\rm P}/\urms(r),
\end{equation}
where $H_{\rm P}$ is the local pressure scale height, changes
substantially between the bottom and the surface. In the Run~A0
$\tau_{\rm to}$ is almost constant in the whole layer whereas in
Runs~B0 and C0 it varies by factors of 9.1 and 31,
respectively. Provided that mixing length arguments hold, the
rotational influence on the flow, measured by the Coriolis number, has
the same variation as a function of radius.

We define the vertical anisotropy parameter as
\begin{equation}
\Lambda_{\rm V} = \frac{2\,\mean{u_r'^2} - \mean{u_\phi'^2} -\mean{u_\theta'^2}}{\urms(r)^2},
\end{equation}
where the averages are taken over $\theta$ and $\phi$, and the primes
denote that $\phi$-averaged mean velocities are subtracted.
For weak stratification, $\Lambda_{\rm V}$ is at most 0.4 in the
middle of the convection zone. Note that $\Lambda_{\rm V}=-1$ at
the boundaries due to the impenetrable boundary conditions. For
Runs~B0 and C0, the turbulence is more anisotropic and the point at
which $\Lambda_{\rm V}$ changes from negative to positive moves closer
to the boundaries. The anisotropy measured by $\Lambda_{\rm V}$ also
increases and in Runs~B0 and C0 it is more than double the value
of Run~A0. In comparison to forced turbulence simulations of
Brandenburg et al.\ (2011a), our results correspond best to cases with
small scale separation, i.e.\ large-scale forcing. This is
consistent with large convective cells that span the whole depth of
the convection zone. The importance of $\Lambda_{\rm V}$ is that in
rotating turbulence, it acts as a source for the vertical
$\Lambda$-effect (e.g.\ R\"udiger 1989; K\"apyl\"a \& Brandenburg
2008) which drives radial differential rotation.

\subsection{Differential rotation}
\label{sec:diffrot}
\begin{figure}
\resizebox{\hsize}{!}
{\includegraphics{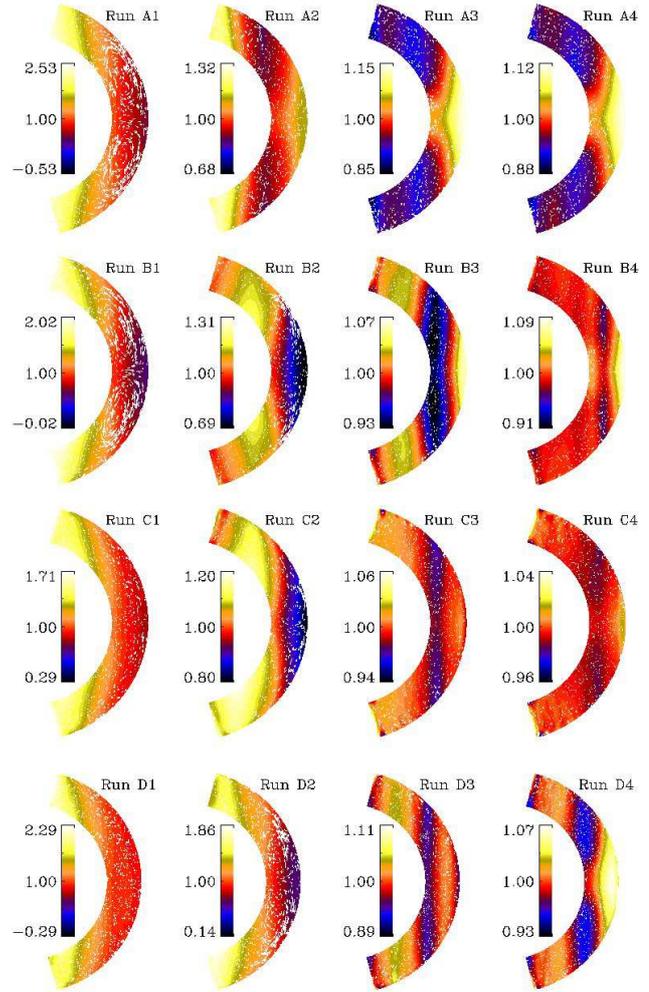}}
\caption{Rotation profiles $\mean{\Omega} = \mean{u}_\phi/(r
  \sin\theta) + \Omega_0$, normalised by $\Omega_0$, from all runs
  with $\Omega_0\neq0$.}
\label{pOm}
\end{figure}

The differential rotation profiles from all runs with $\Omega_0\neq0$ are
shown in Fig.~\ref{pOm}.
We find that an anti-solar differential rotation pattern with strong
meridional circulation forms at the lowest rotation rates. As $\Omega_0$
increases, equatorial acceleration gradually develops. However, in
many cases (e.g.\ Runs~B3, B4, C4, D4) there is a minimum of
the local angular velocity, $\mean{\Omega}=\mean{u}_\phi/r\sin\theta$,
at mid-latitudes and a polar vortex at high latitudes.
Similar profiles have been reported by Miesch et al.\ (2000) and
Elliott et al.\ (2000).
Large-scale vortices arise in Cartesian convection simulations at
sufficiently high rotation rates (Chan 2007; K\"apyl\"a et al.\ 2011b;
Mantere et al.\ 2011). It is unclear whether the polar vortices in
spherical geometry are related to the vortex instability but it is an
intriguing possibility.

\begin{figure}
\resizebox{\hsize}{!}
{\includegraphics{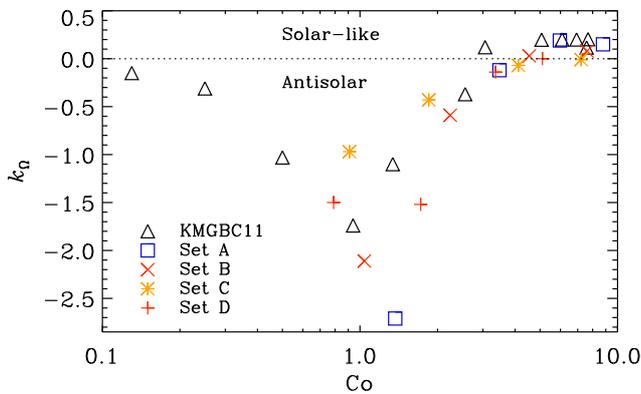}}
\caption{Differential rotation parameter $k_\Omega$ for Sets~A to C as
  indicated by the legend. The black triangles refer to runs taken
  from K\"apyl\"a et al.\ (2011a).}
\label{kOm}
\end{figure}

We quantify the horizontal differential rotation by the parameter
\begin{equation} 
k_\Omega = \frac{\Omega_{\rm p}-\Omega_{\rm eq}}{\Omega_{\rm eq}},
\label{equ:kom}
\end{equation}
where $\Omega_{\rm p} = \onehalf [\Omega(R,\theta_0)+
\Omega(R,\pi-\theta_0)]$, and $\Omega_{\rm eq} =
\Omega(R,\pi/2)$. The results for Sets~A to D, along with
corresponding data from K\"apyl\"a et al.\ (2011a) are shown in
Fig.~\ref{kOm}. We find that for $\Co\approx 1$, $k_\Omega$ is the largest
for the smallest stratification. The results for Sets~B and D seem to
converge for more rapid rotation and produce solar-like rotation ($k_\Omega >
0$) for $\Co > 5$. 

\begin{figure}
\resizebox{\hsize}{!}
{\includegraphics{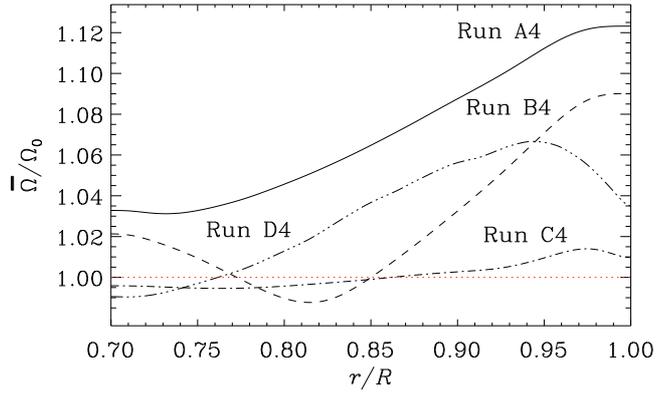}}
\caption{Rotation profiles at the equator from Runs~A4 (solid line),
  B4 (dashed), C4 (dot-dashed), and D4 (triple-dot-dashed).}
\label{pss}
\end{figure}

In Set~C, however, the transition to a solar-like
profile does not yet occur in the parameter range studied here, although
the largest Coriolis number is of the order of 7. 
Another factor that comes into play is the fact that the effective
Reynolds number based on the typical scale of convection cells is
reduced in the runs with the largest stratification. This
seems to be confirmed by the simulations in Set~D, which are the
higher Reynolds number counterparts of the runs in Set~C, although for
Run~D4, $k_\Omega$ is still negative. This is surprising given the
profile seen in Fig.~\ref{pOm} which clearly shows a rapidly rotating
equator. The discrepancy is due to a sharp negative radial gradient of
$\mean\Omega$ near the surface at the equatorial regions in
Run~D4 (see Fig.~\ref{pss}). This is similar to the near-surface shear 
layer observed in the Sun (e.g., Benevolenskaya et al.\ 1999).

The formation of a negative near-surface shear layer only occurs for
strong stratification.
Furthermore, Runs~C4 and D4 (where $\Rey$ is twice as large), suggests
that negative near-surface shear also requires large Reynolds numbers.
It is still unclear whether the current simulations can really capture
the physics of the solar near-surface shear layer (e.g.\ Miesch \& Hindman
2011), but the present results might indicate a path that is worth
following.
Measuring $k_\Omega$ from a little deeper down at $r=0.95R$ gives
$0.06$ which is similar to the results from Runs~B3 and B4 (see
Table~\ref{tab:runs}).

We also note that
the results from K\"apyl\"a et al.\ (2011a) roughly fall in line with
Sets~B and D for $\Co < 3$. However, in the rapid rotation regime, $k_\Omega$
from K\"apyl\"a et al.\ (2011a) is consistently larger than in the
current results with the exception of Set~A. This is probably due to 
the difference in the setups,
i.e.\ we omit here the stably stratified overshoot layer below the
convection zone and the isothermal cooling layer near the surface.

\begin{figure}
\resizebox{\hsize}{!}
{\includegraphics{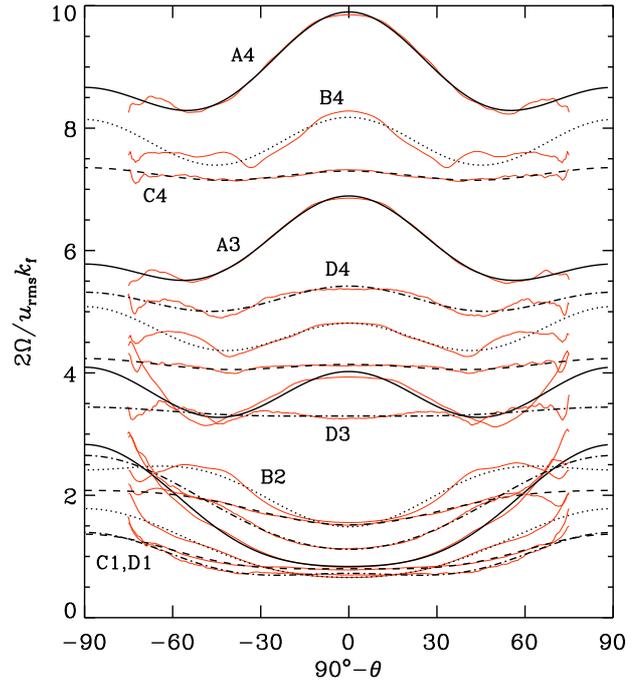}}
\caption{Fits to the latitudinal differential rotation at $r=0.98R$
  for Sets~A (solid), B (dotted), C (dashed), and D (dash-dotted).
  The actual data are shown as thin (red) lines.
}\label{pOmfit}
\end{figure}

It is customary to represent the latitudinal profile of the angular
velocity in terms of Gegenbauer polynomials, i.e.,
\begin{equation} 
\Omega=\Omega_0\sum_{\ell=1,3,5,...}\omega_\ell P_\ell^1(\cos\theta)/\sin\theta.
\label{equ:Omfit}
\end{equation}
(We use here a definition where all associated Legendre polynomials with odd
$m$ values are positive, i.e., $P_1^1(\cos\theta)=\sin\theta$, for example.)
We have determined the coefficients $\omega_1$, $\omega_3$, and $\omega_5$
via a fitting procedure using $\sin^2\!\theta$ as weighting factor to put
more emphasis on the equatorial regions.
The resulting coefficients are given in the last 3 columns of
Table~\ref{tab:runs}.
Note that $\omega_3$ changes sign for $\Co>3$.
This is also consistent with Fig.~\ref{pOmfit}, where we plot the
latitudinal profiles of $2\mean{\Omega}/\urms k_{\rm f}$ together with their
fits from Eq.~(\ref{equ:Omfit}).
The change of sign of $\omega_3$ gives a more robust indicator of the
transition from antisolar to solar-like rotation than just the $k_\Omega$
parameter.

\section{Conclusions}
\label{sec:conclusions}

We study turbulent convection in spherical shells with varying density
stratification. We find that the typical size of convection cells
decreases as the stratification increases, which is in accordance with mixing
length arguments. At the same time the anisotropy of turbulence
increases and the turnover time varies by more than an order of
magnitude from the base to the top of the convection zone.

Although convection seemingly changes greatly as the stratification
increases, the rotation profiles and their qualitative trend as a
function of the Coriolis number change surprisingly little. However,
we find that the results for the largest density stratification are
sensitive to changes of the Reynolds number and that apparently
smaller latitudinal differential rotation for large Coriolis numbers
is obtained for large stratification. Measuring the differential
rotation simply as a difference between the surface values of $\mean{\Omega}$
at the equator and at high latitudes turns out to give misleading
results for our high-resolution runs with the largest
stratification. This is due to the self-consistent generation of a
sharp radial gradient of $\mean{\Omega}$ near the surface, reminiscent of the
near-surface shear layer in the Sun. A similar feature is discernible in
the highly stratified simulations of Bessolaz \& Brun (2011). However,
it remains to be seen whether this is a robust feature.

Another interesting aspect of increasing stratification is related to
the generation of large-scale magnetic fields,
because some of the contributions to
the $\alpha$-effect of mean-field dynamo theory are proportional to
density stratification (e.g.\ Krause \& R\"adler 1980). Furthermore, the
negative magnetic pressure instability (e.g.\ Brandenburg et al.\
2011b; K\"apyl\"a et al.\ 2011c), that can lead to magnetic field
concentrations of the form of active regions, becomes stronger when
stratification increases. We plan to revisit these issues in
forthcoming papers.

\acknowledgements{
  We thank the referee for the suggestion to add the expansion
  in terms of Gegenbauer polynomials.
  Computational resources granted by CSC -- IT Center
  for Science, who are financed by the Ministry of Education, and
  financial support from the Academy of Finland grants No.\ 136189,
  140970 (PJK), 218159 and 141017 (MJM), and the European Research
  Council under the AstroDyn Research Project 227952 (AB) are
  acknowledged.}


\end{document}